\input psfig.sty

\documentstyle[11pt,aaspp4]{article}

\slugcomment{Second Laser Astrophysics Workshop, Submitted to \apjs}

\lefthead{Arnett}
\righthead{Mixing in Astrophysics}

\begin{document}

\title{The Role of Mixing in Astrophysics}

\author{D. Arnett}
\affil{Steward Observatory, University of Arizona, Tucson, AZ 85721}

\begin{abstract}
The role of hydrodynamic mixing in astrophysics is reviewed, emphasizing
connections with laser physics experiments and inertial confinement fusion
(ICF). Computer technology now allows two dimensional (2D) simulations,
with complex microphysics, of stellar hydrodynamics and
evolutionary sequences, and holds the promise for 3D. Careful
validation of astrophysical methods, by laboratory experiment,
by critical comparison of numerical and analytical methods, and
by observation are necessary for the development of simulation
methods with reliable predictive capability. Recent and surprising
results from isotopic patterns in presolar grains, 2D hydrodynamic 
simulations of stellar evolution, and laser tests and computer
simulations of Richtmeyer-Meshkov and Rayleigh-Taylor instabilities 
will be discussed, and related to stellar evolution and supernovae.
\end{abstract}


\section{Introduction}

In astronomy, mixing is important in two widely different situations.
First, there is the mixing of chemically discrete materials. Here we
consider the interstellar medium, and sufficiently cold environments
that solid particles (grains) may survive (\cite{ddc82}). 
This area is exciting now
due to direct experimental identification of presolar grains 
(see \cite{az93,gjw94,ber97}, and references therein).

The second situation involves the mixing of plasma which differs
in its isotopic composition; this is vital to the evolution of stars,
which produce isotopic and nuclear variation by thermonuclear burning
(\cite{ddc68,arn96}). Thermonuclear burning is analogous to chemical
combustion in many ways, and may be as complex. Mixing becomes important
in determining whether flame can spread to new fuel, or is choked by
build up of ashes. Mixing, even in small degree, can provide indications
of ashes which can be used to diagnose burning conditions.

\section{Mixing}

We sometimes forget that stars are really very large. Let us make a 
order of magnitude estimate of diffusion time scales in a dense stellar
plasma. It is the nuclei, not the electrons which define the composition.
The coulomb cross section for pulling ions past each other is of
order $\sigma \approx 10^{-16}\rm\ cm^2$. For a number density
$N \approx 10^{24}\rm\ cm^{-3}$, this implies a mean free path
$\lambda = 1/\sigma N \approx 10^{-8}\rm\ cm$. For a particle
velocity $v_d \approx 10^8\rm\ cm/s$, this gives a diffusion time
$\tau_d = (\Delta r)^2/\lambda v_d \approx  (\Delta r)^2\rm\ s\ cm^{-2}$.
For a linear dimension of stellar size, $\Delta r = 10^{11}\rm\ cm$,
$\tau_d = 3 \times 10^{14}\rm\ y$, or 3,000 Hubble times! While one
may quibble about the exact numbers used, it is clear that pure
diffusion is ineffective for mixing stars, except for extreme cases
involving extremely long time scales and steep gradients.

Actually, we all know from common experience---such as stirring cream
into coffee (tea)---that this discussion is incomplete. To diffusion must be
added {\em advection}, or stirring.
Stars may be stirred too. For example, rotation may induce currents,
as may accretion, and perturbations from a binary companion.
However, the prime mechanism for stirring that is used in stellar
evolutionary calculations is thermally induced convection. 
The idea is that convective motions will stir the heterogeneous
matter, reducing the typical length scale  $\Delta r $ to a value
small enough that diffusion can insure microscopic mixing. 
For our stellar example above, this would require a reduction
in scale of $(\lambda/\Delta r)^{1/2} \approx 10^{-8}$. 
Convection is not perfectly efficient, so that the actual mixing 
time would still be finite. 
Given that such a limit exists, we must examine rapid
evolutionary stages to see if microscopic mixing is a valid 
approximation. For presupernovae, the approximation is almost certainly
not correct, so that these stars are not layered in uniform spherical
shells as conventionally assumed, but heterogeneous in angle as well
as radius.

\section{What The Light Curves Tell Us}

One of the most noticed aspects of SN1987A was the fact that the
progenitor was not a red supergiant, as most stellar evolutionary
calculations predicted, but had a smaller radius, 
$r \approx 3 \times 10^{12}\ \rm cm$. The nature of the HR diagram
for massive stars in the LMC was already an old problem 
(El Eid, et al. 1987, Maeder 1987, Renzini 1987, Truran \& Weiss 1987).
In retrospect, this expectation
of a red supergiant was due to the implicit assumption that semiconvective
mixing was instantaneous, and that the Schwarschild gradient was the one
to use (this is more reasonable for lower mass stars, which
evolve more slowly, see Chapter~7 in Arnett 1996). 
As luck would have it, my formulation of the stellar evolutionary
equations gave the Ledoux criterion as the default, and the progenitor was
a ``blue'' supergiant when the core collapsed (Arnett 1987). 
An example is shown in Figure~\ref{fig1}, with the error box for the observed
progenitor, Sk~-69~202.
This type of behavior is
robust in the sense that, as long as the criterion is similar to the
Ledoux one, LMC star models around 20 solar masses will loop back from the
red giant branch when they do core carbon burning. The actual physical
nature of this mixing process (presumably due to the double diffusion of
ions and heat) does not seem well understood, at least in this context.
The ``blue'' nature of
the progenitor could be due to some other cause, of course, such as
interaction or merger with a binary companion, but no such
complex scenario is required for this position in the HR diagram.

\begin{figure}[t!] 
\psfig{file=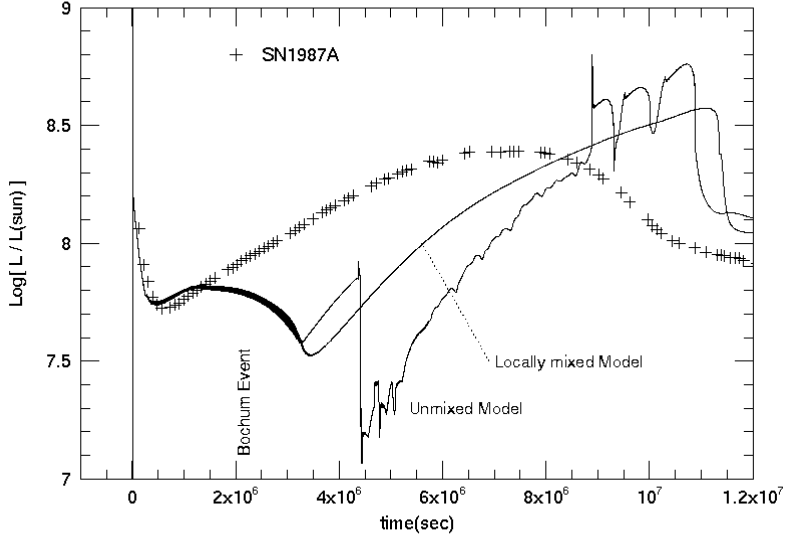,width=4.5in}
\caption{Observed and Simulated Light Curves for SN1987A.}
\label{fig2}
\end{figure}

\begin{figure}

\psfig{file=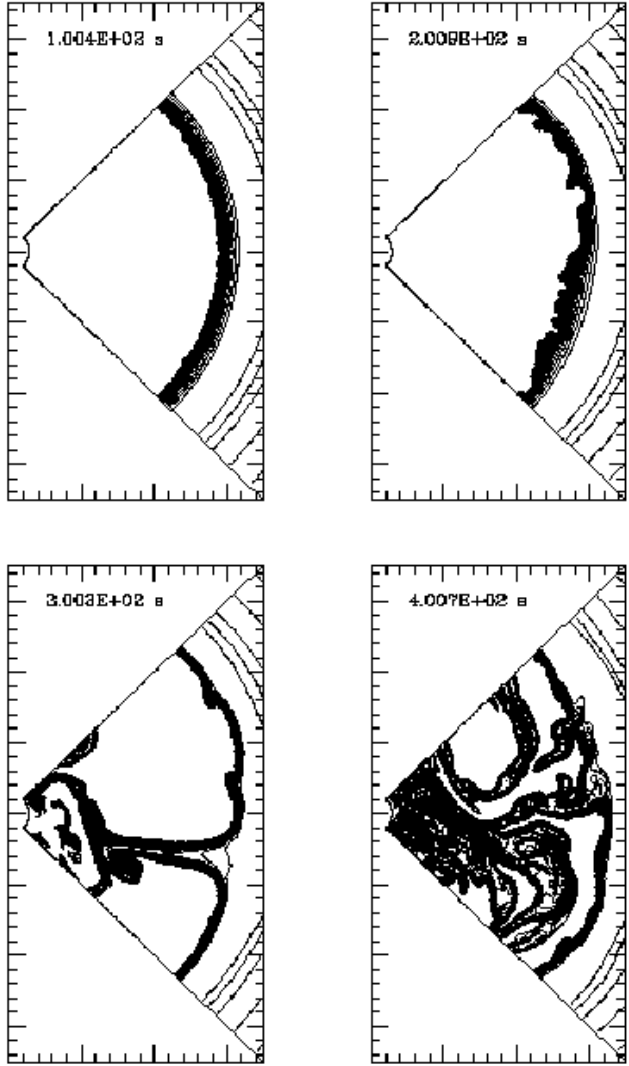,width=5.5in}
\caption{Snapshots of plumes of $\rm^{12}C$, after \cite{ba98}.
Such effects are poorly represented by spherically symmetric simulations.
 \label{fig1}}
\end{figure}

After the euphoria of realizing that SN1987A actually was a close supernova,
I happily simulated the early light curve by running shocks through
stellar models of appropriate radius and mass. This worked fine for the
first twenty days of data (Arnett 1987), but then the agreement degraded
between the observed and computed light curves, as shown in Figure~\ref{fig2}.
Further, synthesis of the spectra was no longer successful at this
epoch (Lucy 1987). This was followed by the Bochum event in the
evolution of the spectra (Hanuschik \& Dachs 1987), 
and later by the early emergence
of x-rays (Donati, et al. 1987, Sunyaev, et al. 1987)
 and detection of $\gamma$-ray lines (Matz et al. 1987). 
Something more was happening than implied by the spherically
symmetric models.

While the powerful tools of radiation hydrodynamics were failing, a
far simpler tool succeeded: after the first two weeks in which the
effects of shock break-out were still felt, an analytic model
(see Arnett 1996 and references therein) reproduced the observed
light curve much more accurately. To obtain the analytic solution,
it was necessary to assume that the opacity was relatively uniform
and the radioactive Ni, while centrally concentrated, was distributed
half-way to the surface! It seemed that some sort of
mixing had occurred after the $\rm ^{56}Ni$ was synthesized in the
explosion.

Various amounts of arbitrary mixing were added to the simulations to
obtain a match to the observed light curve (ABKW 1989).
It is easy to assume mixing as an ad hoc process, but considerably more
demanding to make a plausible simulation of how the mixing actually occurs.
Figure~\ref{fig2} illustrates some options. The observational data for the
UVOIR light curve are shown as crosses. The analytic model fits from
about $t = 10^6\ \rm sec$ onward (see Figure~13.8 in Arnett 1996).
A one-dimensional numerical model without mixing is labeled ``unmixed.''
As the photosphere moves into the burned layers, wild variations in luminosity
occur as the ionization state varies, and hence in the dominant opacity
(which is due to Thomson scattering from free electrons). More careful
treatment of spherical effects will smooth this only slightly (over a time 
$r/c \approx 3 \times 10^{4} \ \rm seconds$). 
If neighboring zones which are convectively
unstable ($\nabla \rho \times \nabla P < 0$) are instantaneously mixed,
the curve labeled ``locally mixed model'' is obtained. This maximal
local mixing case is inadequate to explain the result. This implies that
multidimensional effects---in particular, advection---are in operation.
The mixing is not defined solely by local properties, but rather is
deeply nonlinear.
Subsequent data from much later stages supports this conclusion 
(Wooden 1997, McCray 1997).

\section{Mixing by Advection and/or Diffusion}

Stars are thermonuclear reactors, so that the change of abundances
both drives the evolution and provides a diagnostic of that process.
The rate of change in nuclear abundance $Y_i$ is usually assumed to be
governed by the set of equations,
\begin{equation}
dY_i/dt = - Y_i Y_j R_{ij} - \cdots + Y_k Y_l R_{kl} + \cdots \label{a}
\end{equation}
in which all participating species (denote by indices 
{\it i,\ j\, k,\ l}) are
included. Only binary reactions are explicitly shown, for brevity.
The nuclear reaction rates are denoted $ R_{ij} $, 
with the indices {\it ij} running
through the corresponding species. See Arnett (1996) for detail.

If we are not dealing with homogeneous matter, complications arise 
(Arnett 1997).
First, gradients are not zero, so that we have variations
in both space and time. The ordinary differential equations become
partial differential equations.
As seen from a fixed frame, with material flowing past, the
operator $dY_i/dt$ becomes $\partial Y_i/\partial t 
+ {\bf v\cdot\nabla} Y_i$, where ${\bf v}$ is the fluid velocity.
The new second term is {\em advection}. This gives,
\begin{equation}
\partial Y_i/\partial t 
+ {\bf v\cdot\nabla} Y_i = - Y_i Y_j R_{ij} - \cdots + Y_k Y_l R_{kl} + \cdots, 
\label{b}
\end{equation}
which couples the abundance distribution to the hydrodynamic flow.
Energy release or absorption by nuclear burning further affect the
flow. The system may now be {\em heterogeneous}.

Second, if gradients in composition (e.g., in $Y_i$) are
present, then a new term is generated when we move to the fluid
frame. The velocities of nuclei are split into a symmetric part around the
center of momentum (characterized by a temperature $T$), 
and the fluid velocity $\bf v$. With a composition
gradient, the flux of composition is nonzero, unlike the flux of
momentum in the comoving frame. This gives rise to source terms
due to {\em diffusion} (Landau \& Lifshitz 1959), which in the 
continuity equation for species have the
form $ - {1 \over \rho}{\bf \nabla \cdot F_i }$, where
the composition flux is ${\bf F_i} = \rho Y_i {\bf v}$. Thus,
\begin{equation}
\partial Y_i/\partial t 
+ {\bf v\cdot\nabla} Y_i = - Y_i Y_j R_{ij} - \cdots + Y_k Y_l R_{kl} + \cdots 
 - {1 \over \rho}{\bf \nabla \cdot F_i }. \label{c}
\end{equation}
If the composition gradients are small, we approximate
${\bf F_i} \approx - \rho D {\bf \nabla} Y_i$; this is the usual diffusion
flux for composition, with the diffusion coefficient 
$D \approx \lambda v_d /3$, where  $\lambda $ is the diffusion
mean free path and $v_d $ the mean velocity of diffusing particles
relative to the fluid frame. Notice that the diffusion and advection   
terms may act on strongly differing
length scales in this equation.

We may recover the original simplicity of Eq. \ref{a} if either
(1) the region of interest (the computational ``zone'') is really
homogeneous, or (2) it is very well mixed (large diffusion coefficient D).
In the limit of many mean free paths taken, diffusion approximates a
random walk process. Because of the benign numerical properties of
the diffusion operator, stellar evolutionists have often used
some variety of diffusion to model convective mixing, assuming that
$\lambda$ is approximately a mixing length $\ell$, and many paths were taken.
Note that this involves the singular idea that scales of order 
$\lambda \approx 10^{-8}\rm cm$ are equivalent to
those of order $\ell \approx 10^{+8}\rm cm$ or more.
Ignoring the advection term, this gives,
\begin{equation}
\partial Y_i/\partial t 
 \approx - Y_i Y_j R_{ij} - \cdots + Y_k Y_l R_{kl} + \cdots 
 - {1 \over \rho}{\bf \nabla \cdot F_i }, \label{d}
\end{equation}
which is an approximation to Eq. \ref{c}, and is commonly used
in stellar evolutionary codes (e.g., Woosley \& Weaver 1995). 
It ignores advection, which is the
dominant mode of macroscopic mixing.

\section{Applications to Stellar Hydrodynamics}

In discussion of stellar evolution, one encounters the topics of rotation,
convection, pulsation, mass loss, micro-turbulence, sound waves, shocks,
and instabilities---to name a few---which are all just hydrodynamics.
However, direct simulation of stellar hydrodynamics is limited by causality.
In analogy to light cones in relativity, in hydrodynamics one may define
space-time regions in which communication can occur by the motion of sound
waves. To correctly simulate a wave traveling through a grid, the size of
the time step must be small enough so that sound waves cannot ``jump''
zones. Thus the simulation is restricted to short time steps---an awkward
problem if stellar evolution is desired. While simulations of the solar
convection zone are feasible, the simulation time would be of order hours
instead of the billions of years required for hydrogen burning.
For the latter, a stellar evolution code is used, which damps out
the hydrodynamic motion, obviating the need for the time step 
restriction. Any presumed hydrodynamic motion is then replaced by
an algorithm (such as adiabatic structure and
complete mixing in formally convective regions).
Thus, {\em stellar evolution} deals with the long, slow phenomena,
and {\em stellar hydrodynamics} has dealt with the short term.

However, the stages of evolution prior to a supernova explosion
are fast and eventful. Here direct simulation is 
feasible (\cite{ba98}).
A key region for nucleosynthesis is the oxygen burning shell in
a presupernova star. Besides producing nuclei from Si through Fe prior to
and during the explosive event, it is the site at which the radioactive
$\rm^{56}Ni$ is made and is mixed. The conventional picture of this region
relies upon the notion of thermal balance between nuclear heating and
neutrino cooling in the context of complete microscopic mixing by 
convective motions. 

This is usually treated by the mixing length scenario for convection,
which assumes statistical (well developed) turbulence, random walk of
convective blobs approximated by diffusion, subsonic motions, and almost
adiabatic flow. These approximations are further constrained by a 
simplistic treatment of the boundaries of the convective region.

The time scales for the oxygen burning shell are unusual. The evolutionary time
is $\tau_{evol} \approx 4 \times 10^3\rm\ s$. The convective ``turnover''
time is $\tau_{conv} \approx \Delta r/ v_{conv} \approx \tau_{evol}/10$,
while the sound travel time across the convective region is 
 $\tau_{sound} \approx \Delta r/ v_{sound} \approx  \tau_{conv}/100$.
The burning time is  $\tau_{burn} = E/\varepsilon \le \tau_{conv} $.
Obviously the approximations of subsonic flow, well developed turbulence,
complete microscopic mixing,
and almost adiabatic flow are suspect. 

These time scales are rapid enough to make the
oxygen shell a feasible target for direct numerical simulations, and
an extensive discussion has appeared (\cite{ba98}, and Asida \& Arnett,
in preparation). The two dimensional
simulations show qualitative differences from the previous one dimensional
ones. The oxygen shell is not well mixed, but heterogeneous in coordinates
$\theta$ and $\phi$ as well as $r$. The burning is episodic, localized
in time and space, occurring in flashes rather than as a steady flame.
The burning is strongly coupled to hydrodynamic motion of individual blobs,
but the blobs are more loosely coupled to each other. 

Acoustic and kinetic luminosity are not negligible, contrary to the 
assumptions of mixing length theory. The flow is only mildly subsonic, with 
mach numbers of tens of percent. This gives nonspherical perturbations 
in density and temperature of several percent, especially at the boundaries 
of the convective region.

At the edges of convective regions, the convective motions
couple to gravity waves, giving a slow mixing beyond the formally unstable
region. The convective regions are not so well separated as in the one
dimensional simulations; ``rogue blobs'' cross formally stable regions.
A carbon rich blob became entrained in the oxygen convective shell, and
underwent a violent flash, briefly out-shining the oxygen shell itself 
by a factor of 100. 
Significant variations in neutron excess occur throughout 
the oxygen shell. Because of the localized and episodic burning, the
typical burning conditions are systematically hotter than in one dimensional
simulations, sufficiently so that details of the nucleosynthesis yields
will be affected.

The two dimensional simulations are computationally demanding. Our
radial zoning is comparable to that used in one dimensional simulations,
to which we add several hundred angular zones, giving a computational
demand several hundred times higher.  This
 has limited us to about a quarter of
the final oxygen shell burning in a SN1987A progenitor model. Given the
dramatic differences from one dimensional simulations, it is important
to pursue the evolutionary effects to see exactly how nucleosynthesis
yields, presupernova structures, collapsing core masses, entropies,
and neutron excesses will be changed.  It 
may be that hydrostatic and thermal equilibrium
on average, and the temperature sensitivity of the different burning
stages, taken together, tend to give a rough 
layering in composition, even
if the details of how this happens are quite different.

\section{Toward a Predictive Theory: Tests\\ with 
The NOVA Laser}

Ultimately simulations must be well resolved in three spatial dimensions.
One of the great assets of computers is their ability to represent complex
geometries. If we can implement realistic representations of the
{\em essential} physics, then simulations should become tools to predict---not
``postdict''---phenomena. An essential step toward that goal is the testing
of computer simulations against reality in the form of 
experiment (\cite{re95}).  This is
a venue in which we can alter conditions (unlike astronomical phenomena),
and thereby understand the reasons for particular results. 
Experiments are intrinsically three dimensional, with two dimensional
symmetry available with some effort, so that they provide
a convenient way to assess the effects of dimensionality. 

For Rayleigh-Taylor instabilities, the NOVA experiments not only
sample temperatures similar to those in the helium layer of a
supernova, but hydrodynamically scale to the supernova as well (\cite{ka97}).
In the same sense that aerodynamic wind tunnels have been used in aircraft
design, these high energy density laser experiments allow us to precisely
reproduce a scaled version of part of a supernova.

The NOVA laser is physically imposing. The building is larger in area
than an American football field; the lasers concentrate their beams
on a target about the size a BB (or a small ball bearing). This
enormous change in scale brings home just how high these energy densities
are.
Preliminary results show that the astrophysics code (PROMETHEUS) and
the standard inertial confinement fusion code (CALE) both give
qualitative agreement with the experiment. For example, the velocities
of the spikes and bubbles are both in agreement with experiment,
and analytic theory which is applicable in this experimental
configuration (\cite{ka97}). The two codes give similar, but not identical
results. These differences will require new, more precise experiments
to determine which is most nearly correct.

\clearpage

\acknowledgments

We are grateful to Grant Bazan, Jave Kane, and Bruce Remington for
help and collaboration.

\end{document}